\documentclass[fleqn,usenatbib]{mnras}

\usepackage[T1]{fontenc}

\DeclareRobustCommand{\VAN}[3]{#2}
\let\VANthebibliography\thebibliography
\def\thebibliography{\DeclareRobustCommand{\VAN}[3]{##3}\VANthebibliography}

\usepackage{graphicx}	
\usepackage{amsmath}	
\usepackage{amssymb}	
\usepackage{newtxtext,newtxmath}

\defcitealias{2021arXiv210209568V}{V\&B21}

\title[The merger of clusters/voids in cosmology]{ A possible role for the merger of clusters/voids\\in the cosmological expansion}

\author[S. Mohammadi et al.]{
	S. Mohammadi$^{1}$,\thanks{E-mail: sahar.mohammadi7799@gmail.com}
	E. Yusofi$^{2,3,4}$,\thanks{E-mail: eyusofi@ipm.ir (Corresponding author)}
	M. Mohsenzadeh$^{5}$,\thanks{E-mail: ma.mohsenzadeh@iau.ac.ir}
	M. K. Salem$^{1}$\thanks{E-mail: mkssalem@gmail.com}
	\\
	$^{1}$Plasma Physics Research Center, Science and Research Branch, Islamic Azad University,1477893855, Tehran, Iran\\$^{2}$Department of Physics, Faculty of Basic Sciences, Ayatollah Amoli Branch, Islamic Azad University, Amol, Iran\\$^{3}$School of Astronomy, Institute for Research in Fundamental Sciences(IPM), P. O. Box 19395-5531,Tehran, Iran\\$^{4}$Innovation and Management Research Center, Ayatollah Amoli Branch, Islamic Azad University, Amol, Mazandaran, Iran \\$^{5}$Department of Physics, Faculty of Basic Sciences, Qom Branch, Islamic Azad University, Qom, Iran}

\date{Accepted XXX. Received YYY; in original form ZZZ}

\pubyear{2021}

\begin{document}
	\label{firstpage}
	\pagerange{\pageref{firstpage}--\pageref{lastpage}}
	\maketitle
	
	\begin{abstract}
		In this study, we use the merger process of clusters/voids in the role of variable dark energy fluid to alleviate the Hubble tension, which can lead to a balance in the cosmological expansion rate. To reach this target, we will introduce a modified form of energy density for cosmic fluid with the quadratic equation of state, and then we obtain Hubble, deceleration parameters, and luminosity distance for this fluid. To obtain the merger factor and other parameters of our model, we utilize the \textit{NonlinearModelFit} function within \textit{Mathematica}. By consideration of the local and global measurements of $\rm H_0$, and the equation of state parameter $w$ as the priory values and fitting our model with Observational Hubble Data (OHD) measurements, we will show that the merger of clusters/voids plays the role of balancing the cosmic expansion rate. Also, it will be shown that the model is more compatible than $w$CDM with the standard model to describe the accelerating universe.
	\end{abstract}
	
	\begin{keywords}
		dark energy -- cosmological parameters -- large scale structure of the Universe -- cosmology: theory
	\end{keywords}

	
	
	\section{Introduction}
	\par The current accelerated expansion of the universe has been remaining an unsolved puzzle in cosmology that is supported with extensive observations such as Type Ia SN ~\citep{SupernovaCosmologyProject:1997zqe,SupernovaCosmologyProject:1998vns,SupernovaSearchTeam:1998fmf,Riess:1998dv}, cosmic microwave background (CMB) ~\citep{Planck:2013pxb,Planck:2015mrs}, large scale structure (LSS) and baryon acoustic oscillation (BAO) measurements ~\citep{WMAP:2003elm,SDSS:2005xqv}. There are many theoretical models for interpreting this acceleration, such as the cosmological constant ~\citep{Copeland:2006wr,Carroll:2000fy} that this model is faced with coincidence and fine-tuning problems. Therefore, other alternative models are introduced in order to modify the matter part $V(\phi)$ or geometric side $f(R)$ gravity~\citep{Copeland:2006wr, Carroll:2000fy,Ratra:1988pje,Caldwell:1997ii,Wetterich:1987fm,YOO_2012,Caldwell:2009ix,Bagla:2002yn, Amendola:2006kh, Battye:2015hza}. Also, other suggestions are introduced in different ways, such as Lemaitre-Tolman-Bondi (LTB) metric ~\citep{Lemaitre:1933gd, Moffat;2009mjw} which explain the current acceleration due to existence of inhomogeneity. Recently, our group by introducing a void-dominated cosmology, proposed surface tension of cosmic voids as a possible physical source for dark energy~\citep{Yusofi:2019sai, Yusofi:2022vsg}. 
	\par Until now, the flat $\rm\Lambda $CDM has been more consistent with historical observational data, and the contribution of the cosmological constant $\rm\Lambda $, as a repulsive force originating from vacuum energy, is dominant over the gravitational force resulting from dark matter. But, this model is not able to explain some issues like, such discrepancies in measuring the Hubble constant value $\rm H_{0}$ in local and global observations, which is named $\rm H_0$ tension ~\citep{Kamionkowski:2022pkx, DiValentino:2021izs, Abdalla:2022yfr, Riess:2021jrx, Hu:2023jqc}. Accordingly, Scientists refer to, perhaps there are two reasons for this tension; \\
	$(1)$~The flat $\rm\Lambda CDM $ model needs modification or substitute.\\ 
	$(2)$~Observations are along with imprecise measurements.
	\par Global measurement estimates the value of Hubble constant in the range $ 67-68\quad{\rm{km/s/Mpc}}$, but the higher range of $ 70-75\quad {\rm{km/s/Mpc}}$ are determined directly from distance ladders at local scale ~\citep{Kamionkowski:2022pkx, DiValentino:2021izs, Abdalla:2022yfr}. The most widely supported path for measuring $\rm H_{0}$ in local scale is performed by the SHOES Team ~\citep{DiValentino:2021izs, Abdalla:2022yfr, Riess:2021jrx}. Their recent measurement is $ \rm H_{0} = 73.04 \pm 1.04\quad {\rm{km/s/Mpc}}$, that is resulted by observing Cepheids variables as a distance indicator, that is including systematic uncertainties. Besides this, other projects examine more precise values for $\rm H_{0}$, TRGB project including Chicago-Carnegie Hubble Project (CCHP) determine the value of $ \rm H_{0} $ equal to $ 70 \pm 2\quad {\rm{km/s/Mpc}}$ ~\citep{Freedman:2020dne}. 
	The extragalactic distance database EDD ~\citep{Freedman:2019jwv}, determine this value $ 71.5 \pm 2\quad {\rm{km/s/Mpc}}$ and a measure that calibrates surface brightness fluctuations with TRGB, obtain $ 69.8 \pm 0.8 (\pm 1.1\% \rm \quad stat) \pm  1.7 (\pm 2.4\% \rm \quad sys)\quad {\rm{km/s/Mpc}}$ ~\citep{Freedman:2019jwv, Freedman:2020dne} and the gravitational wave signal, also use for measuring the Hubble parameter ~\citep{Xie:2022brn, Chen:2017rfc}. However, as mentioned earlier, another way for alleviating the Hubble tension is introducing an alternative model for the standard $\rm\Lambda CDM $ model ~\citep{Abdalla:2022yfr}.
	\par  In this paper, $\rm \tilde{V}$CDM model will be studied as a modified form of $\rm\Lambda CDM$ model, that $\rm \tilde{V}$ instead of $\rm\Lambda$, play the role of variable dark energy that earn from the quadratic term in the equation of state which can be originated from the successive merger of voids/clusters at local scales ~\citep{Ananda:2005xp, Ananda:2006gf, Yusofi:2019sai, Pandey:2019qcb, Mos'hafi:2023mcv}. There is some evidence supporting the local void model and its role in cosmic scales ~\citep{Bohringer:2019tyj, DES:2022uvb, Krishnan:2022fzz, Yusofi:2022vsg}. 
	\par For a better understanding of the proposed model, It is necessary to pay attention that the entire visible universe can be thought of as a very large bubble (\textbf{uni}-verse) formed by connecting smaller bubbles (\textbf{multi}-verse) in a hierarchical structure. Therefore, from this point of view, the merger of small voids in the cosmic web can produce a logical connection between local and global scales to alleviate cosmological tensions~\citep{Yusofi:2022vsg, Mos'hafi:2023mcv}. According to this model, in local scales to maintain the second law of thermodynamics, due to the greater gravitational force in the over-dense region of superclusters (at the boundary of the voids) and the lack of gravitational force in the under-dense regions (inside the voids), information, matter, and energy can be transferred from center of supervoids to the center of superclusters ~\citep{Pandey:2019qcb, Pandey:2017tgy}. In this way, the void creates a surface tension for resistance in front being emptier. Thus due to this additional energy and pressure, situated clusters on the border of voids are pushed away from each other. In addition, having this surface energy, the voids can gain the necessary potential to merge with other voids to produce supervoids with lower density and surface tension. Over time, the growth of voids due to successive mergers creates additional pressure on the clusters on voids borders which may act as a possible source of dark energy ~\citep{Yusofi:2019sai, Pandey:2019qcb}.
	\par The main goal of this paper is that if we want to test our model with the aim of reducing the $H_0$ tension, how will the behavior of the merging factor be? We will show that the merger of clusters/voids plays a balancing role between the local and global expansion rate of the universe. The structure of the paper is organized as follows; In section \ref{Sec 2.}, we will briefly explain the complementary role of voids and clusters in the cosmic fluid. We introduce the new energy density resulting from the merger process as a candidate for variable dark energy in section \ref{Sec 3.}. In the next one, we introduce the $\rm \tilde{V}$CDM model to study the dynamics of the universe in comparison with $\rm \Lambda CDM $, and  $w$CDM ~\citep{An:2016keq} models. Then we will obtain Hubble and the deceleration parameters relations in the framework of $\rm \tilde{V}$CDM model. In section \ref{Sec 4.}, we compare the results of the model with other conventional models by fitting them with observational Hubble data. Comparisons will include the deceleration parameter and luminosity distance. Finally, we will bring concluding remarks.
	\section{The Complementary Role of Voids Next to Clusters in the Cosmic Web}
	\label{Sec 2.}
	
	\par The cosmic web is the largest possible structure that is gravitationally formed by the inhomogeneous and anisotropic distribution of matter and describes the transition stage between linear to non-linear structures. The filaments of ordinary matter in the hyper-dense regions of the cosmic web are separated by large, near-spherical regions called cosmic voids, despite being surrounded and concentrated by a halo of cold dark matter. More than 70 percent of the volume of the universe is occupied by cosmic voids and other parts of the cosmic network are filled by superclusters, strings, walls of galaxies, and small voids among them composed of ordinary and dark matter ~\citep{Cautun2014, Correa:2022fmi}. 
	\par It is worth knowing that most theoretical and observational studies are done on the over-dense regions of the cosmic web including small bubbles dominated by filaments and clusters (similar to the void-in-cloud and cloud-in-cloud process). On the other side, the global scale is under-dense including thin walls and shells of clusters that are surrounded by cosmic voids and situated in a void-dominated phase (similar to the cloud-in-void and void-in-void process)~\citep{Redmount:1988dvu, WEYGAERT:2011csd, Sheth:2003py}. This drop-bubble (cluster-void) model is repeated hierarchically in the form of a web-like structure. In this structure, the stars piled up and formed the galaxies, then the galaxies became clusters, and then the clusters merged and formed super-clusters and filaments. Now, if we look from the point of view of void-dominated cosmology ~\citep{Yusofi:2019sai}, it is assumed that at the same time as the above process, the small initial voids also merged together and make bigger bubbles until we reach the largest cosmic voids ~\citep{Lambas:2015afa, Aragon-Calvo:2012, Kirillov:2007ai}. Many observations indicate that the universe is dominated by cosmic voids where dust matter is distributed on their boundary shells/walls ~\citep{Redmount:1988dvu, Aragon-Calvo:2012}. The dynamics of such a cellular and foam-like structure of cosmos produce surface tension (energy) in the boundary layer of voids~\citep{Redmount:1988dvu,vandeWeygaert:2014mqv, Yusofi:2019sai}. The continuous merging of cosmic voids and the simultaneous clustering of the superclusters next to each other increases their possible influence on the universe's formation and evolution ~\citep{Shim:2020wyj}. The increase in the size of cosmic voids after the merger of sub-voids leads to an effective repulsive pressure on the galaxies situated on their shell ~\citep{Yusofi:2022vsg, Redmount:1988dvu, Pandey:2017tgy, Pandey:2019qcb}.
	\par \textbf{Why can the merger of clusters/voids in local scales be important in the cosmic scales?} As we know, the cosmos has a web-like structure on any scale in the hierarchical form~\citep{Redmount:1988dvu, Sheth:2003py}. In local scales the voids (bubbles) are small but as we observe in the larger scales, the main part of the cosmos is the supervoids that are expanding almost in the form of interconnected cells. Also, one of the most important processes that dictate the evolution of voids is their merger in the form of the "void-in-void" process. This process increases the size of the void inside the under-density environment around it. Therefore, it seems that by considering the merger of voids next to clusters in the cosmological models, it is possible to find a way to connect local and global scales. As a result of this connection, it is possible to alleviate some important tensions between theory and observations in astrophysics and cosmology. As confirmation of this hypothesis, we can refer to a recent article published by an international team ~\citep{Farrah:2023opk}. They showed the first observational evidence of the correlation between cosmic expansion and the growth of SMBHs, which implicates an astrophysical source at local scales for dark energy at the cosmic scale. Recently a new method \citep{Benisty:2023vbz} based only on the Local Group dynamics has been developed a method to constrain dark energy from binary galaxies. 
	\section{The Model} 
	\label{Sec 3.}
	\par In this model, we consider that at first, the similar objects merged with each other and finally formed a cluster with a larger structure. For example, first stars merge with each other and form a galaxy, then they merge with each other and form a cluster. Finally, clusters merge with each other and form a supercluster. \\
	Maybe, we can explain the merging of a similar object like a chemical equation that finally obtains an equilibrium constant for them. For example, let's assume that two identical clusters with the number density $n_{\rm clu}$ are merged together and become a supercluster with
	the number density $n_{\rm scl}$ as follow, $$ n_{\rm clu} + n_{\rm clu} \rightarrow n_{\rm scl} $$
	As we mentioned earlier, the voids also can merge together and become a larger supervoid as follow,
	$$ n_{\rm void} + n_{\rm void} \rightarrow n_{\rm svo}~. $$
	Since the superclusters are located on the border of the supervoids, by the expansion of the supervoids the surface tension of them decreases ~\citep{Yusofi:2019sai}. This can potentially result in the easier division of the superclusters into their components, i.e. individual clusters. Therefore, it can be concluded that the enlargement of voids and formation of supervoids on larger scales causes their surface tension to decrease and leads to the division of a supercluster into constituent clusters in the following form:$$ n_{\rm scl}\rightarrow n_{\rm clu} + n_{\rm clu}~. $$ 
	Assuming the simultaneous merger and expansion of voids next to clusters in the cosmic web, we can conclude that the merger process of clusters is in thermal equilibrium as the following form,
	$$ n_{\rm clu} + n_{\rm clu} \rightleftharpoons n_{\rm scl}~.$$
	As a result, we can obtain an \textit{equilibrium constant} $K(T)$ for them as ~\citep{Zemansky:2011hat},
	\begin{equation}
		\label{cvt1}
		K(T)=\frac{n_{\rm scl}}{n_{\rm clu}.n_{\rm clu}}=\frac{n_{\rm scl}}{n^2_{\rm clu}}
	\end{equation}
	Considering ideal gas law $P = nk_{B}T$, and after some straightforward calculations, the pressure relation for the cosmic gas with merger process can be rewritten as follows,
	\begin{equation}
		\label{cvt2}
		P_{\rm scl}=P_{\rm clu}-\frac{P^2_{\rm clu}}{(k_{B}T)}K(T)
	\end{equation}
	As one can see from Equation (\ref{cvt2}), the merger process reduces the primary pressure and acts as a "negative pressure". If we consider the pressure for the gas containing the clusters/voids as
	\begin{equation}
	 P_{\rm clu}=w\rho~, 
	 \end{equation}
	after the merger process, the pressure for the gas containing the produced superclusters/supervoids become as follows, 
	\begin{equation}
		\label{cvt3}
		P_{\rm scl}=w\rho+b\rho^2
	\end{equation}
	In the equation (\ref{cvt3}), $w$ is the equation of state parameter and $ b=\frac {-K(T)w^{2}}{k_{B}T} $, is temperature dependent (of course, in this article, we assume that $b\propto \frac{1}{\rho_{\rm c}}$ as a constant value and $\rho_{\rm c}$ sets the energy scale above which this term becomes relevant ~\citep{Ananda:2006gf}).\\
	The first term of new pressure is $ w\rho$, which can be related to the pressure of galaxies inside of voids (the under-dense regions), and the second one $b\rho^{2}$, can be related to galaxies in the walls on the border of voids with higher density (the over-dense regions)~\citep{Yusofi:2019sai, Ananda:2005xp, Ananda:2006gf, Redmount:1988dvu}. In contrast to galaxy correlations, which are quadratic in the density of galaxies, the cross-correlations of galaxies inside the voids merely depend on their density linearly~\citep{Hamaus:2014afa}. In Fact, our universe consists of fluid with two kinds of phases that coexist together. The under-dense phase is next to the over-dense one. While the under-dense gas is inside the voids, the over-dense phases are situated on the boundary of voids in the form of filaments, clusters, and nodes ~\citep{Redmount:1988dvu, Yusofi:2019sai}. In the proposed model, we consider two types of mergers, the merger of clusters on the one hand and the merger of voids on the other hand.\\
	\subsection{Variable dark energy fluid}
	Now, by considering the pressure for a fluid with merger process (\ref{cvt3}) on \textit{continuity equation} and some straightforward calculations (details are given in ~\citep{Khanpour:2017das}), the density relation, taking into account the merger of clusters/voids is as follows (similar to equation (9) in ~\citep{Ananda:2005xp}),
	\begin{equation}
		\label{rov}
		\rho_{\rm v}= \frac{\rho_{\rm v_0}(1+z)^{3(1+w)}}{1+\alpha(1-(1+z)^{3{(1+w)}})},  
	\end{equation}
	where $\rho_{\rm v_{0}} $ is value of void density in late time and $\alpha=\frac{b\rho_{\rm v_0}}{1+w}$, is called merger factor. Since the parameter $b$ is considered with a negative sign, we will have $ \alpha <0 $, for $ (w > -1) $ and $ \alpha >0 $, for $(w < -1)$. For the special case ($w = -1$), relation (\ref{rov}) replace with the following  model
	\begin{equation}
		\label{rov2}
		\rho_{\rm v}= \frac{\rho_{\rm v_0}}{1-3b\rho_{\rm v_0}\ln{(1+z)}},  
	\end{equation}
	Contrary to the density of the vacuum $\rho_{\rm \Lambda}$, which always has a constant value, the new density defined in our model is always variable with the increase of the scale factor of the universe. So compared to cosmological constant $\Lambda$, we call it $\tilde{\rm V}$. Only in the non-merging model with $b = 0$ is compatible with the cosmological constant. In this respect, it is a generalized form of the standard model. Also, if we consider $\alpha = 0$, the relation (\ref{rov}) returns to the conventional non-merging result \textit{i.e.} $\rho =\rho_{\rm 0} {(1+z)^{3(1+w)}}$.\\
	\subsection{$\tilde{\rm V}$CDM instead of $\Lambda$CDM model}
	In the standard $\Lambda$CDM model, the main contribution of the energy density is divided into two parts $\Omega_{\rm m}$ and $\Omega_{\rm \Lambda}= 1-\Omega_{\rm m}$. In this section, for the first time, we want to withdraw the energy density contribution from the cosmological constant $\Omega_{\rm \Lambda}$ and transfer it to the merger of clusters/voids $\Omega_{\rm v}$.\\
	For the present state of the cosmic dynamics aside from the matter contribution which is agreed on, there is also an important contribution which is attributed to the cosmological constant. We would like here to claim what is affecting the large-scale dynamics is the presence of cosmic voids. We will thus, in this section, deal with and compare the Hubble and deceleration parameters in terms of redshift $z$ for the present $\tilde{\rm V}$CDM model compared with the standard $\rm \Lambda$CDM, and $w$CDM models. From our point of view, the cosmic fluid is a two-phase fluid consisting of two portions; that of matter (ordinary and cold dark matter) denoted by $\Omega_{\rm m}$ and that of cosmic voids denoted by $\Omega_{\rm v}$ ($\equiv \Omega_{\rm \Lambda}$). The basic difference here is that the source for energy is the merger of clusters/voids and is completely known, while the source of cosmological constant \textit{i.e.} $\Omega_\Lambda$ is unknown. It should be noted that by considering the contribution of clusters/voids merger, the basic condition of homogeneity remains intact in our cosmological model. Because on very large cosmic scales we can assume that the universe is made up of nearly identical super voids connected in a web-like structure. Recently, the identicality in density and surface energy of cosmic voids is estimated by heuristic calculations ~\citep{Yusofi:2019sai,Yusofi:2022vsg}.
	\begin{figure}
		\centering
		\includegraphics[width=3.3 in]{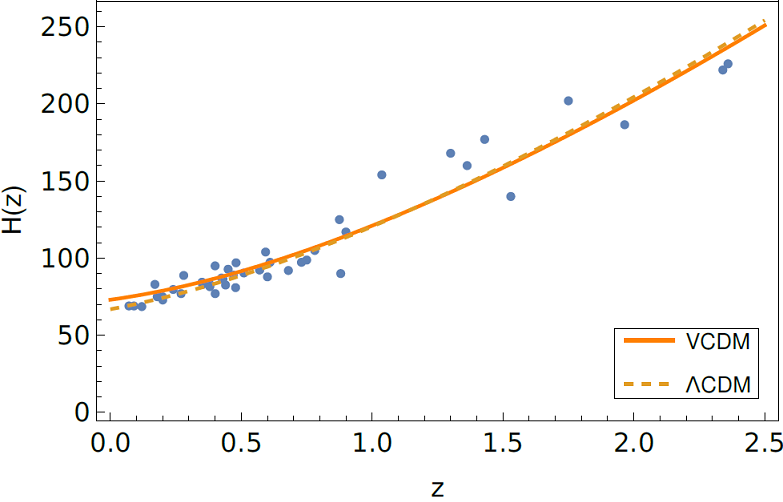}
		\caption{Fitting of $\tilde{\rm V}$CDM (orange) and $\Lambda$CDM (dashed orange) models with the currently available OHD listed in Table \ref{table:I} to determine $\Omega_{\rm m}$, $\Omega_{\rm v}$, and $\alpha$.}
		\label{Fig1}
	\end{figure}
	\begin{table*}
		\caption{The available observational Hubble data \citep{Riess:2021jrx}}
		\centering
		\begin{tabular}{c c c c c c}
			\hline
			\hline
			$z$  & $\rm H(z)$ & Ref. & $z$  & $\rm H(z)$ & Ref.  \\
			\hline
			
			$0.0$&  $67.77 \pm 1.3$ &\citep{DES:2018rjw} & $0.07$&$69.0 \pm 19.6$ &\citep{Stern:2009ep} \\
			$0.09$ & $69.0 \pm 12.0$ & \citep{Simon:2004tf} & $0.01$ & $69.0 \pm 12.0$ & \citep{Stern:2009ep} \\
			$0.12$ & $68.6 \pm 26.2$ & \citep{Stern:2009ep} & $0.17$ & $83.0 \pm 8.0$ & \citep{Stern:2009ep} \\
			$0.179$ & $75.0 \pm 4.0$ & \citep{Moresco:2012jh} & $0.1993$ & $75.0 \pm 5.0$ & \citep{Moresco:2012jh} \\
			$0.2$ & $72.9 \pm 29.6$ & \citep{Stern:2009ep} & $0.24$ & $79.7 \pm 2.7$ & \citep{Gaztanaga:2008xz} \\
			$0.27$ & $77.0 \pm 14.0$ &  \citep{Stern:2009ep} & $0.28$ & $88.8 \pm 36.6$  &\citep{Stern:2009ep} \\
			$0.35$ & $82.7 \pm 8.4$ & \citep{BOSS:2016wmc} & $0.352$ & $83.0 \pm 14.0$ & \citep{Moresco:2012jh} \\
			$0.38$ & $81.5 \pm 1.9$ & \citep{Chuang2013} & $0.3802$ & $83.0 \pm 13.5$ & \citep{BOSS:2016wmc} \\
			$0.4$ & $ 95.0 \pm 17$ &  \citep{Simon:2004tf} & $0.4004$ & $77.0 \pm 10.2$ & \citep{Moresco:2016mzx} \\
			$0.4247$ & $87.1 \pm 11.2$ & \citep{Moresco:2016mzx} & $0.43$ & $86.5 \pm 3.7$ & \citep{Gaztanaga:2008xz} \\
			$0.44$ & $82.6 \pm 7.8$ & \citep{Blake2012} & $0.44497$ & $92.8 \pm 12.9$ &  \citep{Moresco:2016mzx} \\
			$0.47$ & $ 89 \pm 49.6$ & \citep{Ratsimbazafy:2017vga} & $0.4783$ & $80.9 \pm 9.0$ & \citep{Moresco:2016mzx} \\
			$0.48$ & $97.0 \pm 60.0$ & \citep{Stern:2009ep} & $0.51$ & $90.4 \pm 1.9$ & \citep{Chuang2013} \\
			$0.57$ & $96.8 \pm 3.4$ & \citep{Sarmah:2022hmf} & $0.593$ & $104 \pm 13$ & \citep{Moresco:2012jh} \\
			$0.60$ & $87.9 \pm 6.1$ &  \citep{Blake2012} & $0.61$ & $97.3 \pm 2.1$ & \citep{Chuang2013} \\
			$0.68$ & $92.0 \pm 8.0$ &  \citep{Moresco:2012jh} & $0.73$ & $97.3 \pm 7.0$ &  \citep{Blake2012} \\
			$0.781$ & $105 \pm 12.0$ & \citep{Moresco:2012jh} & $0.875$ & $125 \pm 17.0$ & \citep{Moresco:2012jh} \\
			$0.88$ & $90 \pm 40.0$ &  \citep{Stern:2009ep} & $0.9$ & $117 \pm 23.0$ & \citep{Stern:2009ep} \\
			$1.037$ & $154 \pm 20.0$ & \citep{Gaztanaga:2008xz} & $1.3$ & $168 \pm 17$ & \citep{Stern:2009ep} \\
			$1.363$ & $160 \pm 33.6$ & \citep{Moresco:2015cya} & $1.43$ & $177 \pm 18$ & \citep{Stern:2009ep} \\
			$1.53$ & $140 \pm 14$ & \citep{Stern:2009ep} & $1.75$ & $202 \pm 40$ & \citep{Moresco:2015cya} \\
			$1.965$ & $186.5 \pm 50.4$ & \citep{Gaztanaga:2008xz} & $2.3$ & $224 \pm 8.0$ & \citep{Busca2013} \\
			$2.34$ & $222 \pm 7.0$ & \citep{BOSS:2014hwf} & $2.36$ & $226 \pm 8$ & \citep{BOSS:2013igd} \\
			\hline
			\hline		
			
		\end{tabular}
		\label{table:I}
	\end{table*}
	\\
	
	\subsection{Hubble and deceleration parameters in $\rm \tilde{V}$CDM model}
	As is well-known, the equation of state for matter is $w_{\rm m} = 0$. However for $w_{\rm v}$, based on our model (\ref{rov}) we think that its value should be very close to $-1$. Because of the fact that $w_{\rm v}$ is supposed to replace $w_{\rm \Lambda}$, we are allowed to take its value to be very close to $-1$ $(w_{\rm v}\cong -1)$. Fortunately, this assumption is well consistent with the data from Planck 2015 $i.e.$ $w = -1.006\pm 0.045 $~~\citep{Planck:2015fie}. Also combining with Type Ia supernovae (SNe), the dark-energy equation of state parameter is measured to be $w = -1.03\pm 0.03 $~~\citep{Planck:2018vyg}, consistent with our selection.\\
	Recalling that $H$, can be written as
	\begin{equation}
		H^{2}=(\frac{\dot{a}}{a})^2=\frac{8\pi G}{3}\Sigma_{\rm i}\rho_{i}=\frac{8\pi G}{3}(\rho_{\rm m}+\rho_{\rm v})
		\label{eq6},
	\end{equation}
	where the label $i$ includes matter and the vacuum (or void) in our present study. Therefore, instead of the cosmological constant contribution ($\rho_{\Lambda}$) in standard $\rm \Lambda$CDM cosmology, we will use the contribution of merging fluid to produce dark energy and call our proposed model as $\rm \tilde{V}$CDM. Thus, we will obtain the Hubble parameter in terms of $z$ as,
	\begin{equation}
		H_{\rm v}{(z)}=\rm H_{0}\left[\Omega^{0}_{m}{(1+z)}^{3}+\frac{{{\Omega^{0}_{\rm v}(1+z)}^{3(1+w )}}}{1+\alpha\left( 1-{{(1+z)}^{3(1+w )}} \right)}\right]^\frac{1}{2}
		\label{eq7}
	\end{equation}
	for special $\tilde{\rm V}$CDM model in the limit $w \rightarrow -1$, we obtain,
	\begin{equation}
		H_{\rm v}{(z)}=\rm H_{0}\left[\Omega^{0}_{m}{(1+z)}^{3}+\frac{\Omega^{0}_{\rm v}}{1-3b\rho_{\rm v_0}\ln{(1+z)} }\right]^\frac {1}{2}
		\label{eq8}
	\end{equation}
	\begin{figure}
		\centering
		\includegraphics[width=3.3in]{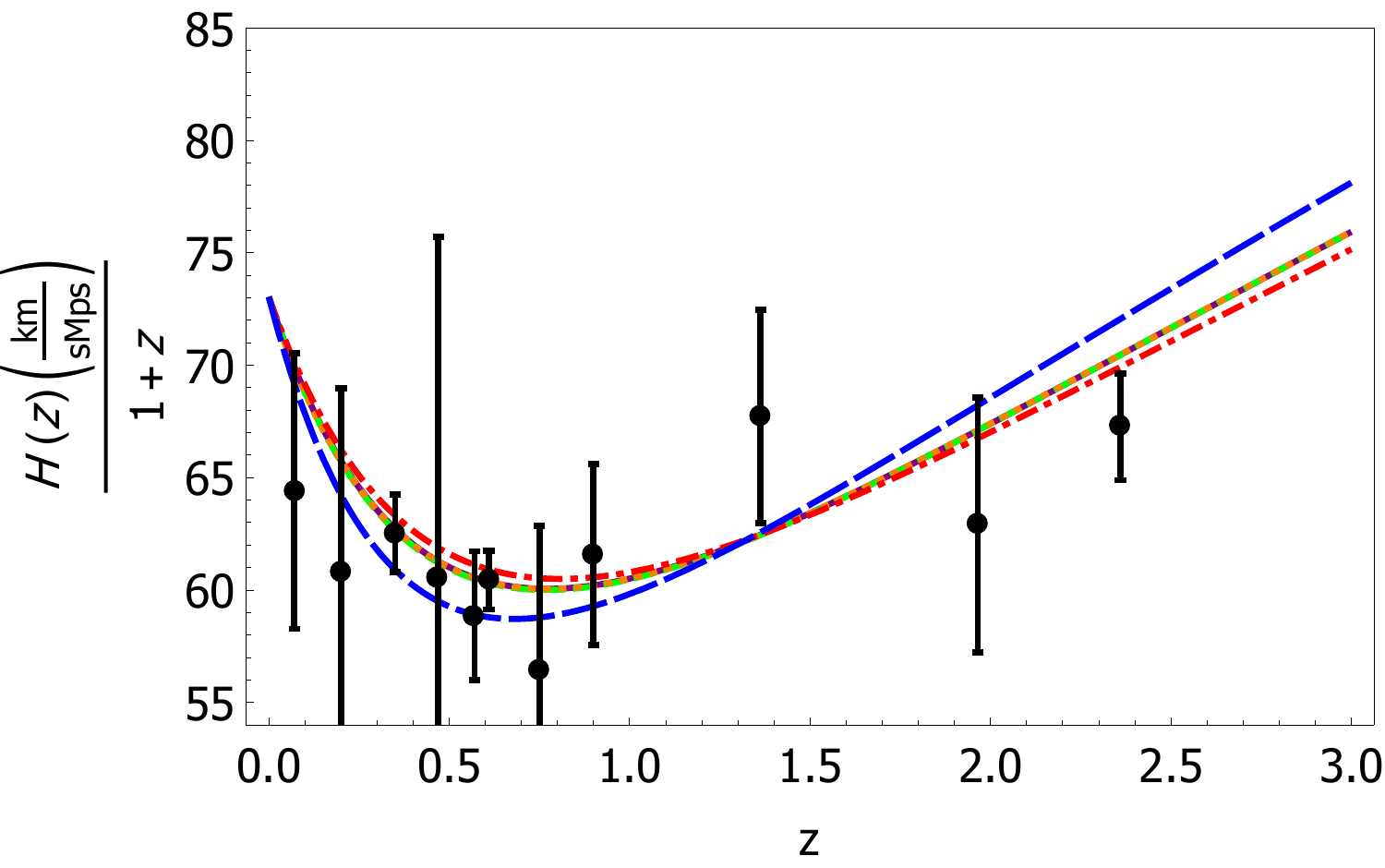}
		\includegraphics[width=3.3in]{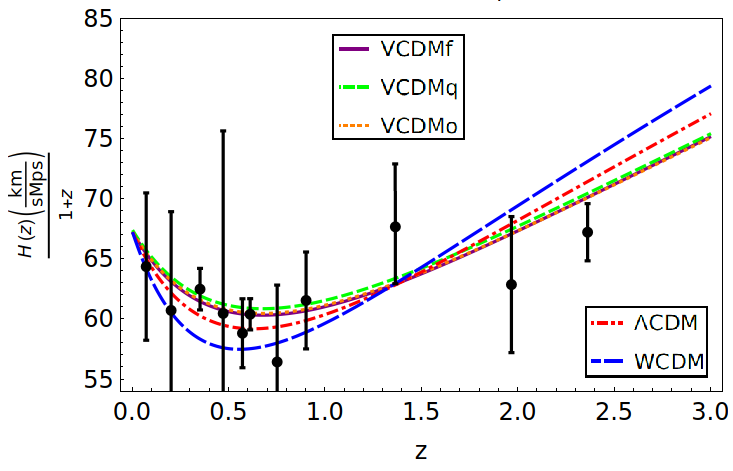}
		\caption{Evolution of rate of scale factor $\left( \frac{H(z)}{1+z}=\dot{a} \right)$ as a function of the redshift, using best-fit cosmological parameters from Table \ref{table:II} (up) and \ref{table:III} (down), for $\tilde{\rm V}$CDM (purple, green and orange),  $\rm\Lambda CDM$ (dashed-dotted red), and $w$CDM (dashed blue). Observational Hubble Data (OHD) points and their error bar are shown in black.}
		\label{Fig2}
	\end{figure} 
	\begin{table}
		\caption{ Cosmological models with prior \textbf{local} Hubble constant, $ H_{0_{L}}$.}
		\centering
		\begin{tabular}{|c| c |c | c |c |c |} 
			\hline\hline
			$\rm Model $ & $\rm H_{0_{prior}}$ & $ w_{v_{prior}} $ & $\Omega_{\rm m}$ & $\Omega_{\rm v} $ & $ \alpha $ \\ [0.5ex] 
			\hline
			$ \rm \tilde{V}CDMf $ &  $ 73.04 $ &  $-1.03$ & $ 0.261 $ & $ 0.739 $ & $ +0.94 $ \\[1ex]  
			$ \rm \tilde{V}CDMo $ &  $ 73.04 $ &  $-1$ & $ 0.261 $ & $ 0.739 $ & $ +0.18 $  \\[1ex]  
			$ \rm \tilde{V}CDMq $ &  $ 73.04 $ &  $-0.97$ & $ 0.261 $ & $ 0.739 $ & $ -3.16 $ \\[1ex]  
			$ \rm\Lambda CDM $&  $ 73.04 $ &  $-1$ & $ 0.253 $ & $ 0.747 $ & $ 0 $ \\[1ex]   
			$ w$CDM & $ 73.04 $ &  $-1.25$ & $ 0.282 $ & $ 0.718 $ & $ 0 $ \\[1ex]  
			\hline
		\end{tabular}
		\label{table:II}
	\end{table}
	The $w$CDM model is like the $ \Lambda$CDM model while we consider the equation of state for this model $ w = -1.25$ ~\citep{An:2016keq}. For the $w$CDM model we obtain
	\begin{equation}
		H_{ w}{(z)}=\rm H_{0}\left[\Omega^{0}_{m}{(1+z)}^{3}+\Omega^{0}_{\rm \Lambda}(1+z)^{3(1+w )}\right]^\frac{1}{2}
		\label{eq9}
	\end{equation}  
	and finally for $\Lambda$CDM model we obtain
	\begin{equation}
		H_{\rm \Lambda }{(z)}=\rm H_{0}\left[\Omega^{0}_{\rm m}{(1+z)}^{3}+\Omega^{0}_{\rm \Lambda}\right]^\frac{1}{2}.
		\label{eq10}
	\end{equation} 
	Also, the deceleration parameter $q$ in cosmology is defined as,
	\begin{equation} \label{hard97}
		q = -\frac{a\ddot{a}}{(\dot{a})^2}=\frac{4\pi G}{3H^2}\Sigma_{\rm i}\rho_{i}(1+3w_{\rm i}).
	\end{equation}
	The larger value of $q$ with a negative sign indicates more rapid acceleration. It is used to quantify the accelerated expansion of the present universe at $t=t_0$. After simple calculations, we will obtain the deceleration parameter in terms of $z$ for $\tilde{\rm V}$CDM model as follows
	\begin{equation} \label{hard909}
		q_{\rm v}(z)=\frac{\Omega_{\rm m}(1+z)^3+\Omega_{\rm v}(1+3w_{\rm v})(\frac{(1+z)^{3(1+w_{\rm v})}}{1+\alpha(1-(1+z)^{3{(1+w_{\rm v})}})})}{2\Omega_{\rm m}(1+z)^3+2\Omega_{\rm v}(\frac{(1+z)^{3(1+w_{\rm v})}}{1+\alpha(1-(1+z)^{3{(1+w_{\rm v})}})})}.
	\end{equation}
	\noindent
	
	\begin{figure}
		\centering
		\includegraphics[width=3in]{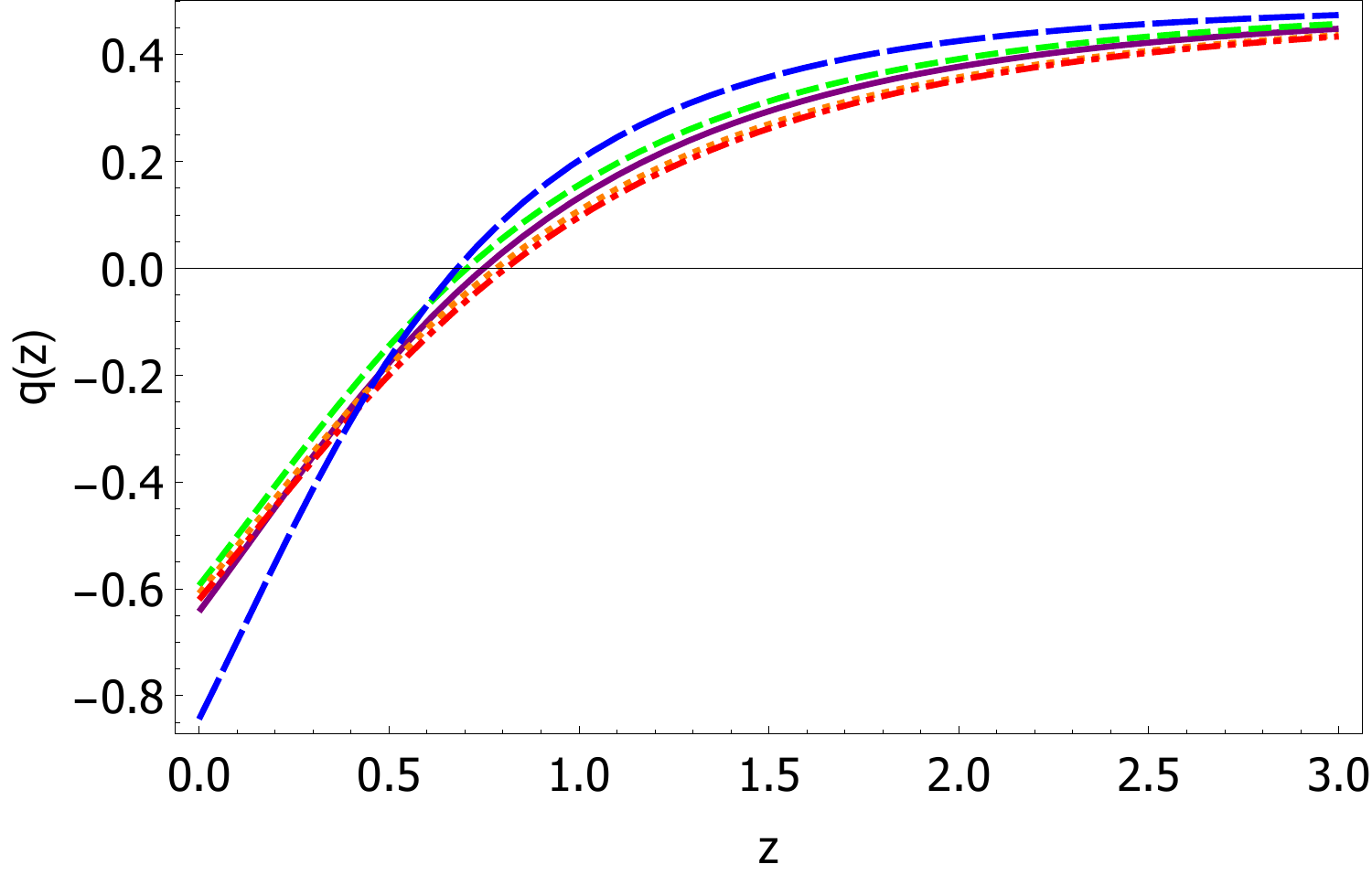}
		\includegraphics[width=3in]{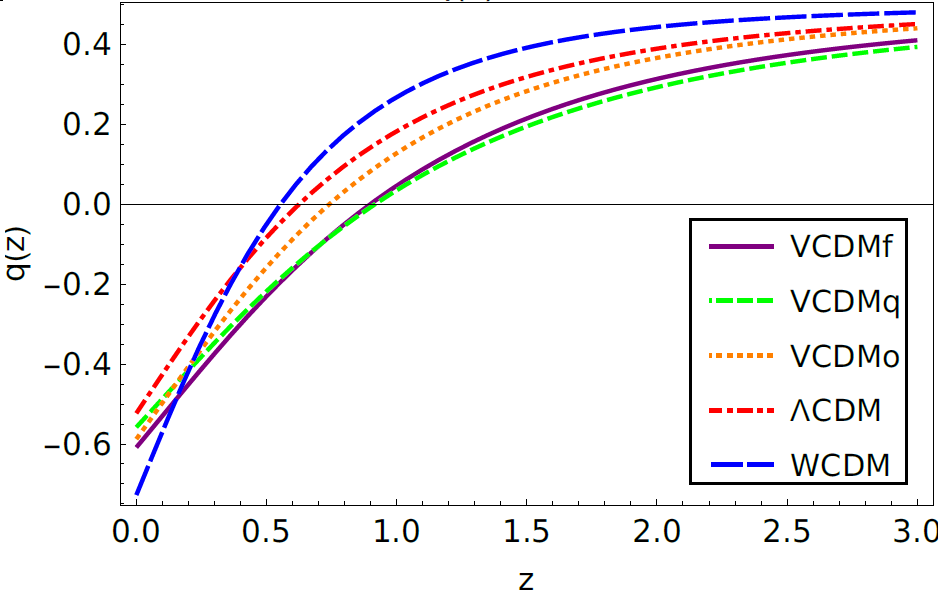}
		\caption{Deceleration parameter $q(z) $ in terms of $z$ based on Table \ref{table:II} (up) and  Table \ref{table:III} (down).}
		\label{Fig3}
	\end{figure}
	\begin{figure}
		\centering
		\includegraphics[width=3in]{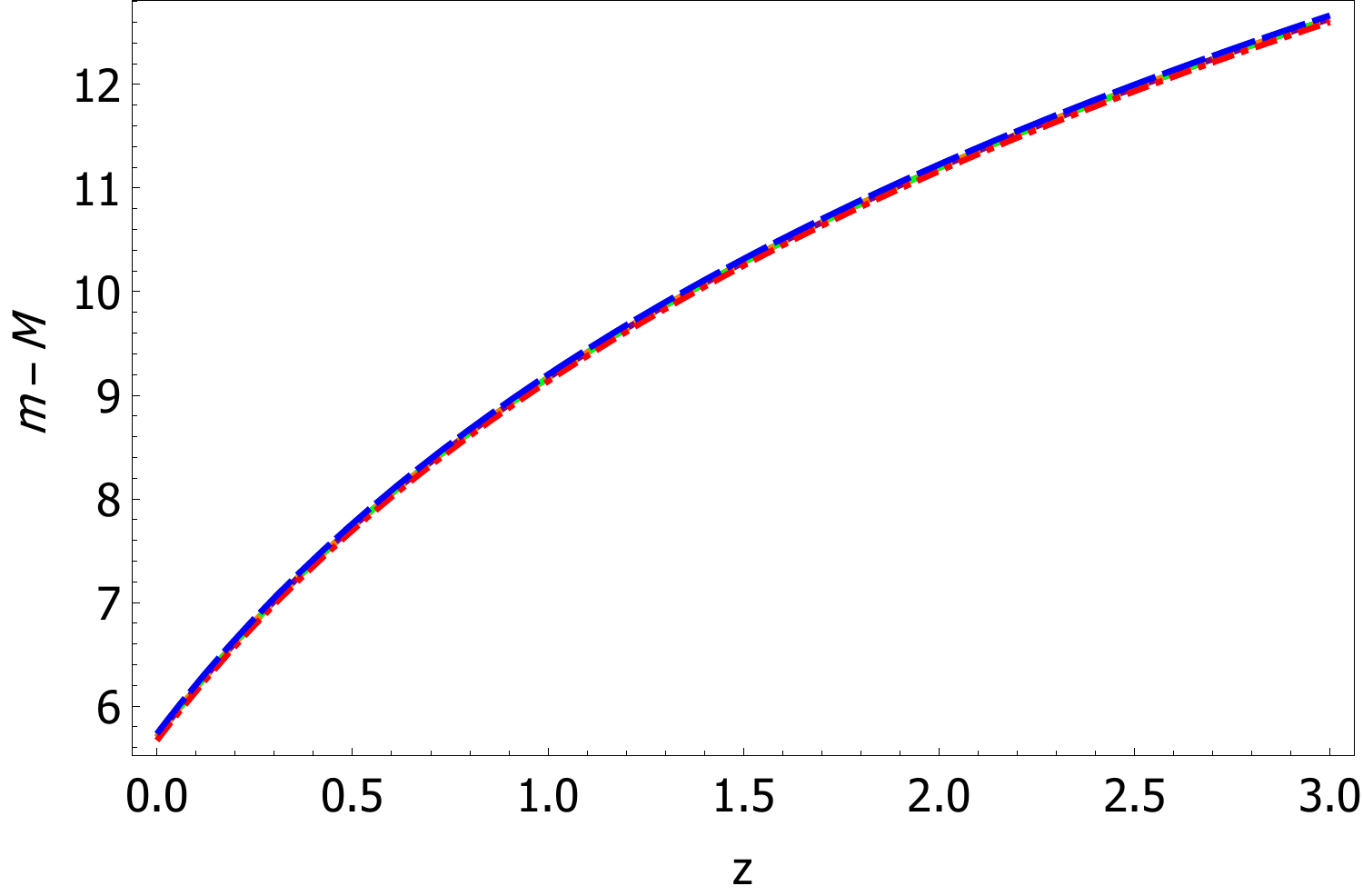}
		\includegraphics[width=3in]{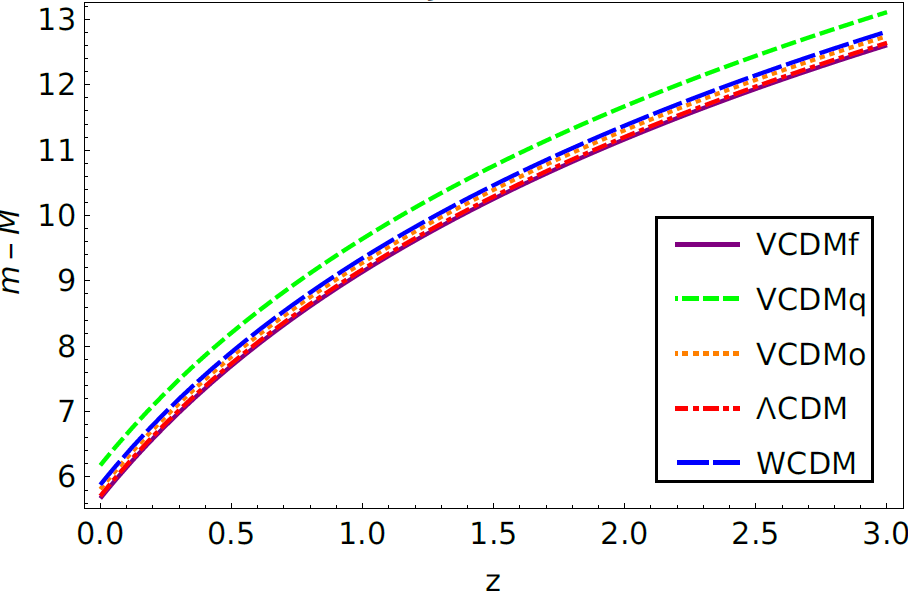}
		\caption{The Luminosity distance parameterized by $m-M$ in terms of $z$ based on Table \ref{table:II} (up) and Table \ref{table:III} (down).}
		\label{Fig4}
	\end{figure}
	\section{Analysis, discussions, and results}
	\label{Sec 4.}
	\subsection{Hubble Parameter}
	To check the models and determine their coefficients, in Fig. \ref{Fig1} we first fit $\rm H(z)$ of different models with $46$ observational available values that are listed in Table \ref{table:I}. For this purpose, we considered the value of $\rm H_0$, and $w_v$ as prior known values to determine $\Omega_{\rm m}$, $\Omega_{\rm v}$, and $\alpha$. We utilize the \textit{NonlinearModelFit} function within \textit{Mathematica} to obtain the merger factor and other necessary parameters for our model. As one can see, the results of this fitting are briefly listed in tables \ref{table:II} and \ref{table:III}. Figures \ref{Fig2}, depicts the Hubble parameter for $\rm \tilde{V}$CDM, $\Lambda$CDM, and $w$CDM models based on tables \ref{table:II} (up) and \ref{table:III} (down). Our model depends on four parameters $\Omega_{\rm m} $, $\Omega_{v}$, $w_v$ and $\alpha$ (merging factor) which $\alpha$ behaves differently in local and global scales.
	\par Fig.~\ref{Fig2} (up), is drawn based on the data in Table \ref{table:II}. Considering the negative sign for $b$ and the three different values of parameter $w_v$, the sign of $\alpha$ works correctly. This type of merger may also be linked to a reduction in acceleration resulting from the merger of galaxies and clusters in over-dense regions. Specifically, if we consider the local measurement of the Hubble constant as a prior value, we have observed a decrease in the Hubble parameter in terms of redshift $z$. As a result, our plot shows a lower value than that of the $\Lambda$CDM model, particularly in values close to zero for $z$.  In this case, the merger factor $\alpha $ can display the merging of the clusters to form superclusters. As a result, the disappearance and collapse of small voids enclosed between them in the local over-dense scales decrease the rate of expansion. Due to clustering and dominance of clusters pressure, similar to the \textit{void-in-cloud} process a small void can not merge with other small voids to occupy more volume and disappear under the pressure of the surrounding clusters. ~\citep{Sheth:2003py}.\\
	\begin{table}
		\caption{Cosmological models with prior \textbf{global} Hubble constant, $ H_{0_{G}}$.}
		\centering
		\begin{tabular}{|c| c |c | c |c |c |} 
			\hline\hline
			$\rm Model $ & $\rm H_{0_{prior}}$ & $ w_{v_{prior}} $ & $\Omega_{\rm m}$ & $\Omega_{\rm v} $ & $ \alpha $ \\ [0.5ex] 
			\hline
			$ \rm \tilde{V}CDMf $ &  $ 67.27 $ &  $-1.03$ & $ 0.277 $ & $ 0.723 $ & $ -6.14 $ \\[1ex]  
			$ \rm \tilde{V}CDMo $ &  $ 67.27 $ &  $-1$ & $ 0.275 $ & $ 0.725 $ & $ -0.50 $  \\[1ex]  
			$ \rm \tilde{V}CDMq $ &  $ 67.27 $ &  $-0.97$ & $ 0.272 $ & $ 0.728 $ & $ +4.95 $ \\[1ex]  
			$ \rm\Lambda CDM $&  $ 67.27 $ &  $-1$ & $ 0.318 $ & $ 0.682 $ & $ 0 $ \\[1ex]   
			$ w$CDM & $ 67.27 $ &  $-1.25$ & $ 0.345 $ & $ 0.655 $ & $ 0 $ \\[1ex]  
			\hline
		\end{tabular}
		\label{table:III}
	\end{table}
	In Fig \ref{Fig2} (down), that is drawn based on the data in  Table \ref{table:III}, the sign of $\alpha $, works the opposite of the previous situation. This means that a positive sign for $b$ can explain how the clusters move away from each other to generate an accelerated expansion. Using the global measurement of Hubble constant $\rm H_{0_{G}} = 67.27$, as a prior value, and the plots of $\rm \tilde{V}$CDM models show the higher rates of expansion compared to the standard $\Lambda$CDM model, while the wCDM model behaves similarly in both cases. The error bars in Fig. \ref{Fig2} are related to the currently available OHD measurements listed in Table \ref{table:I}. According to the void-dominated universe, energy (information) must be exchanged between clusters and voids due to the second law of thermodynamics~\citep{Pandey:2017tgy,Pandey:2019qcb}. In this way, the center of the clusters on the border of voids becomes denser and the center of the cosmic voids becomes more empty. Due to the increase in the volume of the voids after merging, the surface tension between the clusters located at the border of the voids decreases and with the reduction of surface tension at the boundary of the supervoids, the clusters move away from each other more easily, and probably this process accelerates their expansion at larger scales ~\citep{Hoffman:2017ako,Yusofi:2019sai}. As the voids become the dominant volume of our universe, they gain enough potential to merge with other voids to create larger supervoids similar to \textit{void-in-void} process~\citep{Sheth:2003py}. Expansion of the supervoids can exert negative pressure and anti-gravitational force on superclusters at cosmic scale~\citep{Yusofi:2019sai, Pandey:2019qcb,Hoffman:2017ako}.
	\subsection{Deceleration parameter}
	
	Fig. \ref{Fig3} shows the ability of $\rm \tilde{V}$CDM model to explain accelerated expansion in global scales. 
	Using the $H_{0_{G}}$ as priory value, we obtain larger acceleration than $\rm\Lambda$CDM model for $\rm \tilde{V}$CDM, that is due to the existence merger of vast voids at global scales. As we can see in the upper plot in Fig. \ref{Fig3}, due to the dominance of gravity between the clusters and the disappearance of small voids between them in local scales, the gravitational attraction overcomes the repulsion in such scales, and as a result, the acceleration in the local era of the universe decreases. But in the bottom plot, due to the dominance of the merging of vast voids, the acceleration of the universe in global scales obtains larger values compared to the standard $\rm\Lambda$CDM model. At the global scale, the merger of vast voids that form the super cosmic voids with lower surface energy causes the rate of expansion to increase. Further, in contrast to $w$CDM, in both cases, the $\rm \tilde{V}$CDM models have more consistency with the $\rm\Lambda CDM$ model.
	\subsection{Luminosity distance}
	Fig. \ref{Fig4}, shows the increasing rate of $m-M$ based on global data from Table \ref{table:III}. As a result, based on these plots, it can be concluded, supernovas inside the border of voids would be observed to be receding more rapidly than expected, as compared to other supernovas inside the void \cite{Moffat;2009mjw}.
	\begin{equation}
		d_{L}=\frac{(1+z)}{\rm H_{0}}\int^{3} _{0}\frac{dz}{E(z)}
		\label{eq20}
	\end{equation}
	\begin{equation}
		m-M=5Logd_{L}+25
		\label{eq21}
	\end{equation}
	Hear $E(z)=\frac{H(z)}{\rm H_{0}}$. By using local data, all plots behave similarly to each other, but by global data based on Table \ref{table:III}, the $\rm \tilde{V}$CDM model predicts a bigger luminosity distance than $\rm \Lambda$CDM model. On the other hand, for this model, the brightness of apparent magnitude $m$ appears much dimmer than absolute magnitude $M$, which can be a sign of the expansion of the universe due to the merger of vast voids at the larger scales.
	\section{conclusions}
	\label{Sec 5.}
	We have introduced a variable energy density, $\rho_{\rm v}$, for the cosmic fluid in our research. The variable dark energy fluid includes the merger process of clusters/voids instead of $\rm \rho_{\rm \Lambda}$ and has the potential to alleviate the $\rm H_0$ tension while balancing the cosmological expansion rate. Additionally, we have obtained a modified form for the Hubble and deceleration parameters, as well as the luminosity distance, by adding the merger term in the energy density relation. Our study has shown that the merging fluid plays two opposite roles in balancing the cosmic expansion rate and is also more compatible with the standard model and OHD measurements. Notably, the merger of clusters/voids in the $\Tilde{\rm V}$CDM model acts as an accelerator in expansion rate and behaves as a de Sitter-like space at the cosmic scale, but as a decelerator at local scales.\\
	By choosing the local value of $\rm H_0$ as a priory value, the merger of clusters causes a decrease in the expansion rate. By choosing the global value of $\rm H_0$ as a priory value, the merger of vast voids causes an increase in the expansion rate. Even more interesting is that both of these scenarios result in reducing the Hubble tension.
	\\
	Furthermore, our statistical tests, including the Chi-square test, have been completed and will be detailed in a separate article. The results clearly demonstrate the reduction of Hubble tension. We believe our model falls within the categories of both interacting dark energy and late universe models. In the future, we plan to explore the physical and cosmological implications of the model and its possible connection with the merger process in high-density regions, such as supermassive black holes at local scales.
	
	\section*{Data availability}
	The data underlying this article will be shared on reasonable request to the corresponding author.
	\section*{Acknowledgements}
	First, we would like to our appreciation to the anonymous referees for providing valuable feedback that significantly contributed to the quality of our paper's presentation. Also, we want to thank A. Talebian for his contribution to the project, including his insightful comments and questions about the model and his assistance with creating the plots and tables. EY would like to thank H. Firouzjahi, M. Khanpour, B. Khanpour, H. Moshafi, and M. Ramzanpour for their constructive input during our discussions and for helping to establish accurate hypotheses for the model. This work has been supported by the Islamic Azad University, Science and Research Branch, Tehran, Iran.
	
	\bibliographystyle{mnras}
	\bibliography{mnras_sahar_R2} 
\end{document}